\documentclass[12pt]{article}
\usepackage[top=1.0in, bottom=1.0in, left=0.75in, right=0.75in]{geometry}
\usepackage{amsmath, amsthm, amssymb}
\usepackage[sorting=nyt,citestyle=numeric,bibstyle=numeric,backend=bibtex,maxbibnames=99]{biblatex}
\allowdisplaybreaks

\usepackage{caption}
\usepackage{amsmath}
\usepackage{algorithm}
\usepackage[noend]{algpseudocode}
\usepackage{graphicx}
\usepackage{subfig}
\usepackage{booktabs}
\usepackage[affil-it]{authblk}
\usepackage{graphics}
\usepackage{mathrsfs}
\usepackage{setspace}
\usepackage{epstopdf}
\usepackage{rotating}
\usepackage{longtable}
\usepackage{tabularx}
\usepackage{tabulary}
\usepackage{ltxtable}
\usepackage{multirow}
\usepackage{listings}

\newsavebox\affbox
\title{\textbf{Adversarial Machine Learning in Network Intrusion Detection Systems}}

\author[1]{Elie Alhajjar%
	\thanks{Email: \texttt{elie.alhajjar@westpoint.edu}; Corresponding author.}} 
\author[1]{Paul Maxwell%
    \thanks{Email: \texttt{paul.maxwell@westpoint.edu}.}}
\author[1]{Nathaniel D. Bastian%
    \thanks{Email: \texttt{nathaniel.bastian@westpoint.edu}.}}

\affil[1]{Army Cyber Institute, U.S. Military Academy\\West Point, New York 10996}

\date{}

\begin{document}

\maketitle

\abstract{Adversarial examples are inputs to a machine learning system intentionally crafted by an attacker to fool the model into producing an incorrect output. These examples have achieved a great deal of success in several domains such as image recognition, speech recognition and spam detection. In this paper, we study the nature of the adversarial problem in Network Intrusion Detection Systems (NIDS). We focus on the attack perspective, which includes techniques to generate adversarial examples capable of evading a variety of machine learning models. More specifically, we explore the use of evolutionary computation (particle swarm optimization and genetic algorithm) and deep learning (generative adversarial networks) as tools for adversarial example generation. To assess the performance of these algorithms in evading a NIDS, we apply them to two publicly available data sets, namely the NSL-KDD and UNSW-NB15, and we contrast them to a baseline perturbation method: Monte Carlo simulation. The results show that our adversarial example generation techniques cause high misclassification rates in eleven different machine learning models, along with a voting classifier. Our work highlights the vulnerability of machine learning based NIDS in the face of adversarial perturbation.\\
	
}

\newpage


\section{Introduction}

It is becoming evident each and every day that machine learning algorithms are achieving impressive results in domains in which it is hard to specify a set of rules for their procedures. Examples of this phenomenon include industries like finance \cite{WONG1997301, BOSE2001211}, transportation \cite{10.1162/neco.1991.3.1.88}, education \cite{stanczyk2007machine, 10.1561/1500000005}, health care \cite{KOUROU20158} and tasks like image recognition \cite{simonyan2014deep, He_2015, He_2016}, machine translation \cite{sutskever2014sequence, Chen_2018}, and speech recognition \cite{vaswani2017attention, Lee_2018, Zhang_2018, xiong2016achieving}.

Motivated by the ease of adoption and the increased availability of affordable computational power (especially cloud computing services), machine learning algorithms are being explored in almost every commercial application and are offering great promise for the future of automation. Facing such a vast adoption across multiple disciplines, some of their weaknesses are exposed and sometimes exploited by malicious actors. For example, a common challenge to these algorithms is ``generalization'' or ``robustness'', which is the ability of the algorithm to maintain performance whenever dealing with data coming from a different distribution with which it was trained. 

For a long period of time, the sole focus of machine learning researchers was improving the performance of machine learning systems (true positive rate, accuracy, etc.). Nowadays, the robustness of these systems can no longer be ignored; many of them have been shown to be highly vulnerable to intentional adversarial attacks. This fact renders them inadequate for real-world applications, especially mission-critical ones. Adversarial machine learning is classified by that National Institute of Standards and Technology (NIST) into four categories of attacks: evasion, extraction, poisoning, and inference.  In this work, we examine adversarial examples, which is a type of evasion attacks.

An adversarial example is an input to a machine learning model that an attacker has intentionally designed to cause the model to make a mistake. In general, the attacker may have no access to the architecture of the machine learning system being attacked, which is known as a black-box attack.  Attackers can approximate a white-box attack by using the notion of ``transferability'', which means that an input designed to confuse a certain machine learning model is able to trigger a similar behavior within a different model.  In this work, we model a white-box attack by evaluating our examples against a variety of machine learning models, thus showing performance over a wide range of possible systems.

Network Intrusion Detection Systems (NIDS) monitor network traffic to detect abnormal activities, such as attacks against hosts or servers. Machine learning algorithms offer the benefit of detecting novel differences in network traffic through training on normal and attack traffic. The traditional approach to designing a NIDS relies on an expert human analyst codifying rules that define normal behavior and intrusions. Due to the frequent failures of this approach to detect novel intrusions and the desire to lower the analyst's workload, machine learning models are incorporated into NIDS with the goal of automating the process and supplementing the human effort.

On the other hand, the adoption of machine learning algorithms in the network intrusion detection setting has raised a major security issue, namely the introduction of adversarial machine learning. In this work, we study adversarial machine learning through the lens of NIDS to show the sensitivity of machine learning models when faced with adversarial attacks generated by perturbation techniques based upon evolutionary computation, deep learning, and Monte Carlo methods.

In precise terms, we use three different perturbation methods to generate adversarial examples: particle swarm optimization (PSO), a genetic algorithm (GA), and a generative adversarial network (GAN). We apply these algorithms to two well-known data sets, NSL-KDD and UNSW-NB15, and show that the perturbations created by these techniques are able to fool eleven different machine learning models, plus an ensemble model. For our baseline adversarial example generation method, we use a Monte Carlo (MC) simulation for random generation of perturbations.

The paper is organized as follows. In Section \ref{sec:Pre}, we give an overview of the field of machine learning. We set the notation and establish the state of the art taxonomy therein. In Section \ref{sec:Lit}, we review the most relevant and up to date literature that deals with adversarial machine learning. Section \ref{sec:Meth} describes the methodology used to generate adversarial examples. We explain the intuition as well as the technical details of our chosen perturbation methods. In Section \ref{sec:Res}, we apply our evolutionary computation and deep learning methods to the two data sets mentioned above, and we record the results while contrasting them with the baseline perturbation method. We also discuss our findings and distinguish between the machine learning models that are highly sensitive and the ones that show some type of robustness. We conclude our work in Section \ref{sec:Conc} and pose some open questions and future research directions. 

\section{Prerequisites}\label{sec:Pre}

Machine learning encompasses a vast field of techniques that extract patterns from data, as well as the theory and analysis relating to these algorithms. A computer program is said to learn from experience $E$ with respect to some class of tasks $T$ and performance measure $P$, if its performance at tasks in $T$, as measured by $P$, improves with experience $E$ \cite{mitchell1997machine}.

In this section, we review the basic concepts of machine learning, broadly divided into three major areas: supervised learning, unsupervised learning, and reinforcement learning. We then introduce the notion of adversarial machine learning and categorize the types of attacks along three dimensions: timing, information, and goals. We adopt a simplistic approach for the sake of exposition, the reader interested in the technical definitions and details is referred to the books \cite{joseph2019adversarial,vorobeychik2018adversarial, mitchell1997machine}.

\subsection{Supervised Learning}

Consider a data set $D=\{(x_i,y_i):i=1,\dots,n\}$ which lies in a product space $Z=X \times Y$, where $\times$ denotes the cartesian product. We call $X$ the input space or feature space and $Y$ the label space or the output space. In general, $D$ is assumed to be drawn from an unknown distribution.

For a model class $F$ of functions mapping the input space $X$ to the output space $Y$, the ultimate goal is to find a model $f \in F$ which is a good approximation of the labels actually observed in the data. Given a labeled training data set $D$ and a cost function $l$, the process goes as follows: the cost function assigns a numerical value to each combination of data point, true label, and classifier label. The classifier then tries to ``learn'' the function that minimizes the cost incurred in predicting the label $y_i$ of a new data point $(x_i, y_i)$ when only the predictor variable $x_i$ is revealed to the classifier.

In simple terms, if there is some ``true” function $h$ we are trying to learn, the goal is to find a function $f \in F$ which is as ``close'' to $h$ as possible, given the constraints that the model class imposes on us. When the labels are real numbers ($Y=\mathbb{R}$), the paradigm is called regression learning, and when $Y$ is a finite set of labels it is called classification learning.

In regression learning, since the label values are real numbers, a loss function is used to penalize predictions that are made far from the observed labels. An appropriate set of functions is the set of $l_p$-norms defined as:
\begin{equation}
    ||f(x)-y||_p := \Big(\sum_{i=1}^{n}|f(x_i)-y_i|^p\Big)^\frac{1}{p},
\end{equation}
where $x \in X$ is the input value, $y \in Y$ is the true label and $f(x)$ is the value predicted by the model $f$. The most commonly used norms are the $l_1$-norm and the $l_2$-norm, also known as the Manhattan norm and the Euclidean norm, respectively. A well-known example here is linear regression, in which the model class $F$ is the set of all linear functions and the standard method is the least squares method that minimizes the $l_2$-norm.

In classification learning, the labels form a finite set of values. For our purposes, we consider a binary set of labels $Y=\{0,1\}$ where $``0"$ indicates a benign network traffic and $``1"$ indicates a malicious network traffic. The most natural loss function in this setting is a function that takes the value $``1"$ if the predicted label matches the real label and $``0"$ otherwise. Due to its non-convexity, this function is sometimes replaced by the hinge loss function or the logistic loss function.

\subsection{Unsupervised Learning}

In unsupervised learning, a typical data set consists only of features without labels, that is, \\$D=\{(x_i):i=1,\dots,n\}$. Hence, problems in this paradigm are concerned with identifying aspects of the joint distribution of observed features rather than predicting a target label. Common techniques include clustering, principal component analysis (PCA), and matrix completion.

In clustering, the data set $D$ is partitioned into a collection $S$ of subsets such that each subset of $S$ contains feature vectors that are close to each other based on a chosen metric or distance. Examples of clustering are the k-means clustering and the Gaussian mixture model.  Challenges with this method is determining the numbers of clusters and which clusters are indicative of malicious behavior.

In principal component analysis, the feature vectors $x_i$ of the data set $D$ form the rows of a matrix $A$ of dimension say $m$. The task becomes to find a collection of orthonormal basis vectors of $A$ of size $k<m$. These vectors are the eigenvectors of the matrix $A$ and they correspond to the columns of a basis matrix $B$. Then any feature vector $x_i$ can be reconstructed as $x_i'=VV^Tx_i$, where $T$ denotes the transpose of a matrix. The error can be calculated:
\begin{equation}
    Error_i=x_i-x_i'=(I-VV^T)x_i,
\end{equation}
where $I$ is the $m \times m$ identity matrix, and the goal is eventually to minimize such error in order to get a good approximation of the original data.

In matrix completion, the general goal is to recover a complete matrix from a few given observations. It is usually impossible to complete an arbitrary matrix with only partial observations, so additional assumptions are needed to proceed. A crucial assumption for example is that the matrix has rank much smaller than the number of its rows and columns. In this context, the Frobenius norm is the most commonly used norm to minimize the error incurred by the matrix completion method.

\subsection{Reinforcement Learning}

In reinforcement learning, the goal is to learn via interaction and feedback, or in other words learning to solve a task by trial-and-error. The mathematical foundation for reinforcement learning is based on Markov Decision Processes (MDPs), which we define below.

A discrete-time MDP is a tuple $(S,A,T,r,\delta)$ where $S$ is a finite set of states, $A$ is a finite set of actions, $T$ is the set of transition probabilities, $r$ is a reward function, and $\delta \in (0,1)$ is a discount factor. Loosely speaking, reinforcement learning aims at maximizing the reward function for a state $s \in S$ and an action $a \in A$. Two techniques are widely used in this framework: $Q$-learning and policy learning.

In $Q$-learning, a function $Q$ is defined that takes as input one state and one action and returns the expected reward of that action and all subsequent actions at that state. Before learning begins, $Q$ is initialized to a random fixed value and it gets updated as the learning proceeds until it converges to an optimal value.

In policy learning, a mapping $\pi: S \rightarrow A$ is introduced from the set of states to the set of actions. We refer to this map as policy: $a=\pi(s)$ means that when state $s$ is observed, the best thing to do is to take action $a$. The task then becomes to find a policy with maximum expected return.

\subsection{Adversarial Machine Learning}

Adversarial Machine Learning (AML) is a research field that lies at the intersection of machine learning and computer security. Here, we present a general categorization of attacks in the context of machine learning. Below we classify possible attacks into three main pillars.

First, the timing of the attack plays a crucial role in the process. The main distinction occurs between evasion attacks and poisoning attacks. The former ones are executed at decision time: they assume that the model has already been trained, so the attacker aims at changing its behavior to cause the model to make incorrect predictions. The latter ones, in contrast, take place before the model is trained. They aim at modifying a part of the data used for training purposes to corrupt the final model.

Second, we highlight the nature of information the attacker has about the learning algorithm. This allows us to compare white-box attacks and black-box attacks. Namely, white-box attacks assume that the model is known to the adversary, whereas in black-box attacks the adversary has limited or no information about the model, although may obtain some of the information indirectly, for example, through queries and captured responses.

Third, attackers may have different reasons for attacking, such as evading detection or reducing confidence in the algorithm. We differentiate between two broad classes of attack goals: targeted attacks and reliability attacks. In a targeted attack, the attacker’s goal is to cause a misclassification into a specific label or target. In contrast, a reliability attack aims to degrade the perceived reliability of the machine learning system by maximizing prediction error without any particular target label.

\section{Related Work}\label{sec:Lit}

The first publication that hints at adversarial machine learning dates back to the early $2000$'s. In $2004$, Dalvi et al. \cite{10.1145/1014052.1014066} managed to fool classifiers for spam detection by adding intentional changes to the body of an email. In $2005$, Lowd and Meek \cite{Lowd_2005} introduced the adversarial classifier reverse engineering (ACRE) learning problem, the task of learning sufficient information about a classifier to construct adversarial attacks. They presented efficient algorithms for reverse engineering linear classifiers with either continuous or Boolean features and demonstrated their effectiveness using real data from the domain of spam filtering.

A taxonomy for attacks and defenses in adversarial setting was first established by Barreno et al. in \cite{Barreno_2006} and later refined in \cite{Barreno_2010}. This taxonomy clearly defines threat models and includes the concept of adversarial examples without any mentioning of the term. The first time this term was coined was in $2013$ when Szegedy et al. \cite{szegedy2013intriguing} investigated properties of deep neural networks and found that by applying a certain imperceptible perturbation to an image, these networks misclassified it with high confidence. Moreover, they showed that the specific nature of these perturbations is not a random artifact of learning: the same perturbation can cause a different network, that was trained on a different subset of the data set, to misclassify the same input.

The last decade has witnessed a large body of publications in the realm of adversarial machine learning in several domains and it is impossible to summarize it all in a paragraph or two. We point the curious reader to a couple of excellent recent articles that attempt to survey the field of AML. Liu et al. \cite{Liu_2018} investigate security threats and give a systematic survey on them from two aspects, the training phase and the testing/inferring phase. They also categorize existing defensive techniques of machine learning into four groups: security assessment mechanisms, countermeasures in the training phase, those in the testing or inferring phase, data security, and privacy. Further, Akhtar and Mian \cite{Akhtar_2018} focus on adversarial attacks on deep learning in computer vision. They emphasize that adversarial attacks are possible in practical conditions and review the contributions that evaluate such attacks in real-world scenarios.

A more detailed survey is provided by Serban et al. in \cite{alex2018adversarial}. The authors introduce an exhaustive list of attacks/defenses and thoroughly discuss the property of transferability. They provide a taxonomy of attacks and defenses that is specific to the adversarial examples field. This taxonomy is meant to properly structure the methods in AML and help future researchers to position their work in comparison to other publications.

Most of the adversarial machine learning research so far has focused on unconstrained domains (e.g., image and object recognition). This is mainly due to the following assumption: the adversary is always able to fully exploit each feature or pixel of a given image. Several attack algorithms currently exist in the literature and many variations to their models are widely implemented. These models include the Fast Gradient Sign Method (FGSM) \cite{goodfellow2014explaining}, the Jacobian-based saliency map attack (JSMA) \cite{Papernot_2016}, Deepfool \cite{Moosavi_Dezfooli_2016}, and the Carlini Wagner attack (CW) \cite{Carlini_2017}.

Goodfellow et al. \cite{goodfellow2014explaining} proposed the FGSM to generate adversarial perturbations based on the gradient of the loss function relative to the input image and, thus, enable computational efficiency through backpropagation. Papernot et al. \cite{Papernot_2016} created adversarial saliency maps by computing forward derivatives, which are used to identify the input feature to be perturbed towards the target class. Moosavi-Dezfooli et al. \cite{Moosavi_Dezfooli_2016} proposed an approach to find the closest distance from original input to the decision boundary of adversarial examples. Carlini and Wagner \cite{Carlini_2017} introduced three new gradient-based attack algorithms ($l_2$, $l_\infty$, and $l_0$) that are more effective than all previously known methods in terms of the adversarial success rates achieved with minimal perturbation amounts. 

Unlike unconstrained domains, the situation is quite different in constrained ones due to the following three characteristics:
\begin{enumerate}
    \item The values within a single feature can be binary, continuous or categorical.
    \item The values of different features in a data set can be correlated with one another.
    \item Some features are fixed and cannot be controlled by a potential adversary.
\end{enumerate}

For these reasons, it was not clear whether constrained domains are less vulnerable to adversarial example generation than unconstrained domains. In \cite{sheatsley_adversarial_nodate}, 
Sheatsley tested the above hypothesis by creating targeted universal perturbation vectors that encode feature saliency within the envelope of domain constraints. The experiment was able to generate misclassification rates greater than $95 \%$ by introducing two algorithms: the adaptive JSMA, which crafts adversarial examples that obey domain constraints, and the histogram sketch generation, which produces adversarial sketches.

In this work, we aim to provide more evidence to the same hypothesis. We adapt the algorithms used in GA, PSO, and GAN to generate adversarial examples in unconstrained domains. By doing so, we achieve high misclassification rates in the majority of commonly used machine learning models applied to the data sets NSL-KDD and UNSW-NB15. Our baseline MC simulation perturbation method generates more or less random perturbations in the data.

We end this section by mentioning that similar techniques appeared previously in the adversarial machine learning literature \cite{shah2019evaluating}. Hu and Tan \cite{hu2017generating} propose a GAN-based algorithm named MalGAN to generate adversarial malware examples which are able to bypass black box machine learning detection models. Nguyen, Yosinski and Clune \cite{Nguyen_2015} use evolutionary algorithms to generate images that are given high prediction scores by convolutional neural networks. Drawing ideas from genetic programming, Xu, Qi, and Evans \cite{xu2016automatically} propose a generic method to evaluate the robustness of classifiers under attack. Their key idea is to stochastically manipulate a malicious sample to find a variant that preserves the malicious behavior but is classified as benign by the classifier. More recently, Alzantot et al. \cite{10.1145/3321707.3321749} introduced GenAttack, a gradient-free optimization technique that uses genetic algorithms for synthesizing adversarial examples in the black box setting. Mosli et al. \cite{mosli2019giants} created AdversarialPSO, a black-box attack that uses fewer queries to create adversarial examples with high success rates. AdversarialPSO is based on particle swarm optimization, and is flexible in balancing the number of queries submitted to the target compared to the quality of imperceptible adversarial examples. Finally, Devine and Bastian \cite{DevineBastian} used a machine learning approach for robust malware classification that integrates a MC simulation for adversarial perturbation with meta learning using a stacked ensemble-based methodology.

\section{Methodology}\label{sec:Meth} 

In this section, we give a technical description of the features in the data sets used in our experiments. We then explain the details of the techniques employed for adversarial example generation. This leads to the layout of our computational setting.

\subsection{Data sets}

There are a small number of publicly available, well-known labeled data sets of network traffic for cyber security research.  Two of these data sets are: NSL-KDD \cite{10.5555/1736481.1736489} and UNSW-NB15 \cite{7348942}.  Both data sets have a mix of benign and malicious traffic with a variety of network traffic types and attack types.  Both data sets have quality limitations due to factors like generation methods, prevalence of attack traffic, and size. Nevertheless, these data sets are commonly used to evaluate machine learning based NIDS.

In general, the data sets contain fields with raw and processed data derived from the underlying network traffic.  Normally, this data is engineered in research work to detect intrusions in a network.  In many works \cite{grosse_adversarial_2016, foley_adversarial_2017,elsayed_adversarial_2018}, effort is made to alter the data to create adversarial examples that can fool detection systems and classifiers.  In domains such as image and speech recognition, perturbing the data to fool the systems can be done in a manner that does not affect the functionality or appearance of the original item.  For example, changing pixel values in an image by a few bits can be undetectable to the human eye yet fool a trained machine learning classifier.  In network security applications, which is an unconstrained domain, care must be taken when generating adversarial examples that alter the source data.  There are data, such as the duration of a network packet flow, that if altered will not impact the functionality of the protocol.  However, there are fields that if changed can cause the functionality of the transmission to fail, such as changing the protocol type from TCP to UDP.

The goal of an attacker (who may use adversarial examples) is to alter packets to achieve a desired effect while still maintaining the functionality of the traffic.  Previous research \cite{grosse_adversarial_2016,foley_adversarial_2017} acknowledge that certain fields in the data sets are immutable while others are open to perturbations.  In this work, we subscribe to this position and place technical constraints on data fields that are mutable. For the data sets used, the mutable fields are described below.

\subsubsection{NSL-KDD Data Set}

This data set is an altered version of the KDD’99 data set with over 125,000 training samples and more than 22,000 test samples. As described in \cite{iglesias_analysis_2015}, it contains 41 features based upon raw network traffic elements, flow traffic elements, and content. The data contains attacks from four attack types that were more prevalent at the time of the set's creation (Denial of Service, Probes, Remote-to-local, and User-to-Root). Attack traffic comprises a little over 50 percent of the data in the set and the normal data in the set is not representative of modern traffic. Engineering the features using common techniques such as one-hot encoding and Min-Max Scaling results in a final data set with 121 features. Of these features, some are immutable such as \textit{protocol-type}, \textit{service}, and \textit{flag}. A change to these values would cause the underlying traffic to become non-functional. The remaining features are mutable, though we limit the perturbations to changes that increase the initial values. The rationale behind such decision is that these fields, such as $src\_bytes$, can increase without altering the functionality of the traffic (e.g. null packets can be inserted). Decreasing these fields without expert knowledge of the traffic content would not be certifiably feasible. Other constraints on perturbations to the data are fields that are binary (the value can only be flipped from $0$ to $1$ and vice versa) and fields that are linearly related such as $same\_srv\_rate$ and $diff\_srv\_rate$ must sum to 1. A change to one of these fields alters the content of the other. Our analysis of the engineered features results in 93 immutable features and 29 mutable ones.

\subsubsection{UNSW-NB15 Data Set}

The UNSW-NB15 data set was generated at the Cyber Range Lab of the Australian Centre for Cyber Security. It contains over 175,000 training samples and 82,000+ test samples that are labeled as benign or malicious (including attack type). The data contains nine types of more modern, low footprint attacks as described in \cite{7348942}. The benign traffic is more reflective of current network traffic as well. In this set, attacks compromise a much smaller percentage of the data at under 25 percent of the total. This data set was also purposely constructed to more closely match the content of the training set and test set as highlighted in \cite{doi:10.1080/19393555.2015.1125974}. The data has entries with 49 fields of which some are generated by network flow analyzers and intrusion detection tools.  Engineering these fields results in 196 features of which we identified 23 as mutable. Similar to the NSL-KDD data set, some fields are constrained as binary, increasing in value only, or have a linear relationship.  However, some fields in this set can have decreasing values such as $sttl$ (e.g. the time to live can arbitrarily be set by the sender), and $sload$ (e.g. the source can increase/decrease packet sizes and transmission rates).

\subsection{Adversarial Example Generation}

Adversarial examples have been extensively used to evade machine learning systems. The methods of generation for these adversarial examples include mostly heuristic techniques \cite{foley_adversarial_2017,sheatsley_adversarial_nodate,yang_generative_2017} and GANs \cite{goodfellow_generative_nodate,lin_idsgan:_2018,hu2017generating}. In this work, we introduce two novel techniques for adversarial example generation that use evolutionary computation, PSO and GA, along with a GAN that uniquely considers constrained feature sets to avoid the failure of the resulting traffic. Finally, we implement a MC simulation generator for baseline comparison.

\subsubsection{Genetic Algorithm}

\begin{figure}
  \centering
  \subfloat[Initial Chromosome transformation into operable (mutable) and fixed (immutable) parts. Cells marked with an ‘F’ are immutable and those with an ‘M’ are mutable.]{\includegraphics[width=0.3\textwidth]{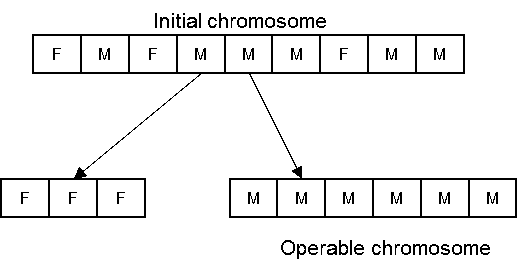}\label{fig:f1}}
  \hfill
  \subfloat[Example cross-over operation on two parent operable chromosomes and the resulting child operable chromosomes.] {\includegraphics[width=0.3\textwidth]{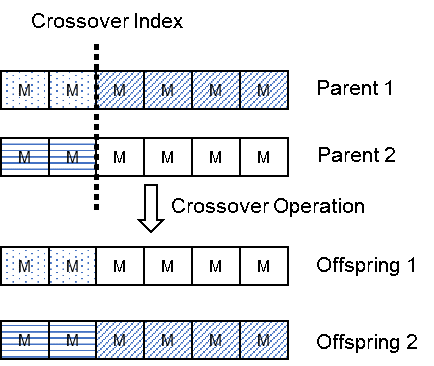}\label{fig:f2}}
  \hfill
  \subfloat[Example mutation operation performed on an operable chromosome.] {\includegraphics[width=0.3\textwidth]{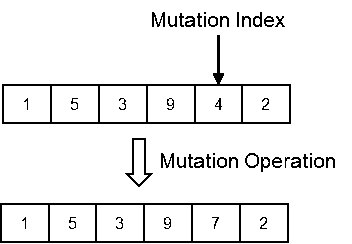}\label{fig:f3}}
  \caption{Genetic Algorithm Operators.}
  \label{fig:GA}
\end{figure}

Genetic algorithms are commonly used to find near-optimal solutions to NP-hard problems \cite{whitley1994genetic}. In the NIDS setting, searching the space with over 90 dimensions for a solution that best fools a machine learning classifier can benefit from this type of search heuristic. Figure \ref{fig:GA} illustrates a simplistic view of how a genetic algorithm works in general.

Genetic algorithms typically operate on data structures known as chromosomes; a chromosome is a representation of the problem’s data. For this work, a chromosome is comprised of a data feature from the source data sets. The GA operates on each row entry in the data set creating a population of chromosomes. This initial population contains a seed chromosome based on the original row’s values and randomly generated chromosomes based upon the seed. Each chromosome contains both mutable and immutable elements. As a result, prior to performing the algorithm’s standard operations (cross-over and mutation), the chromosome is separated into two parts, the fixed and operable elements as shown in Figure \ref{fig:f1}. Of note, each chromosome in the population has a common, fixed element. Once the chromosomes are sub-divided, the heuristic operations can be performed on the operable sub-part of the chromosome.

Based upon the chosen cross-over rate, the cross-over operation is performed. Two parent chromosomes are randomly selected from the population and a cross-over index is selected randomly. This index separates the operable chromosome into left and right portions in each parent. The right portions are then swapped between parents to form two new operable chromosomes. These child chromosomes are then inserted into the population. The cross-over operation is illustrated in Figure \ref{fig:f2}.

The mutation operation perturbs values of randomly selected cells based on a chosen mutation rate. The operation first randomly selects chromosomes to operate upon based on the mutation rate. Then, a mutation index is randomly selected. Using this index, the targeted chromosome is modified as shown in the example of Figure \ref{fig:f3}. The resulting chromosome is left in the population in its modified form.

Upon completion of the cross-over and mutation operators, the chromosomes’ fitness functions are evaluated. In this domain, the fitness metric is the output probability that the chromosome is classified as benign by a machine learning classifier. To evaluate the chromosome, the operable chromosome must be recombined with the immutable part to form a valid chromosome. The best chromosome for each generation is maintained in subsequent generations. The chromosomes in the next generation, limited by the chosen population size variable, are selected using a roulette wheel technique. The process of selection, cross-over, and mutation is repeated until a chosen number of iterations is reached or the improvement in the best fitness is below a chosen threshold. A pseudo-code for the GA is shown in Algorithm \ref{alg:GA}.

\begin{algorithm}
\caption{Genetic Algorithm}\label{alg:GA}
\begin{algorithmic}[1]
\State Choose cross-over rate, mutation rate, number of chromosomes, maximum iterations
\State Separate seed vector into immutable and mutable sub-vectors
\State Generate initial population of mutable vectors
\State\hspace{\algorithmicindent} Mutable seed vector created as a chromosome
\State\hspace{\algorithmicindent} Randomly generate remaining chromosomes
\State\hspace{\algorithmicindent} Determine population best fitness using fixed portion of seed vector
\While {(iterations $<$ maximum iterations) and (improvement $>$ improvement minimum)}
    \For {number of cross-over operations}
        \State Randomly select two parent chromosomes
        \State Randomly select cross-over index
        \State Perform cross-over at the index
        \State Add two offspring chromosomes to the population
    \EndFor
    \For {each chromosome in the population}
        \If {random number $<=$ mutation rate}
            \State Randomly select mutation index
            \State Update chromosome at mutation index with new value
        \EndIf
    \EndFor
    \State Calculate fitness of each chromosome
    \State Update population best fitness
    \State Select next generation using roulette wheel method
\EndWhile
\end{algorithmic}
\end{algorithm}

\subsubsection{Particle Swarm Optimization}

\begin{figure}
  \centering
  \subfloat[Example Particle Swarm with particles moving in the search space.] {\includegraphics[width=0.5\textwidth]{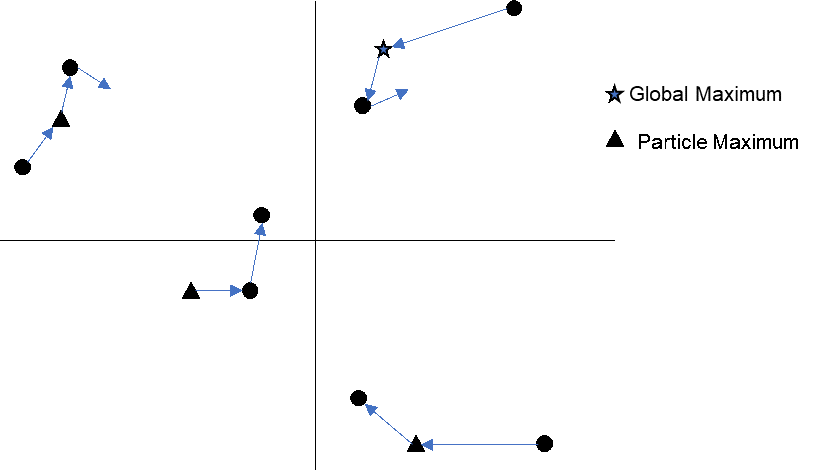}\label{fig:ff1}}
  \hfill
  \subfloat[Example Particle after creation of the particle swarm.] {\includegraphics[width=0.4\textwidth]{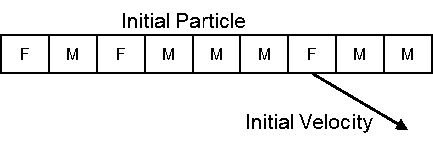}\label{fig:ff2}}
  \caption{Particle Swarm Optimization Scheme.}
  \label{fig:PSO}
\end{figure}

Particle swarm optimization is a bio-inspired heuristic based on behaviors of animals that flock or swarm \cite{clerc2010particle}. The idea is that each member or particle of the swarm explores the search space with a calculated velocity. The velocity is updated based upon the particle’s best solution and the swarm’s optimal solution. The probability that the velocity is influenced by the best solutions is determined by selected weighting coefficients. We used the PySwarms code base \cite{james_v._miranda_pyswarms:_2018} to develop the heuristic used in our work. Figure \ref{fig:ff1} illustrates how the PSO algorithm works.

The heuristic begins with the creation of the swarm. The swarm is seeded with a row from the evaluated data set. The remaining particles are then randomly created using the seed particle as a baseline. Each particle is then randomly assigned an initial velocity. Similar to the GA, the particle is composed of mutable and immutable cells. An example particle is shown in Figure \ref{fig:ff2}.

In each iteration of the PSO algorithm, the particles’ fitness function is evaluated using the same metric as the GA. The values of the particle’s best location and the global best location are updated as required. Next, the distances between each particle’s location and the particle’s best location and the global best location is determined. These values are then used to update the particle’s velocity and location. The heuristic is repeated until either a selected number of iterations is reached or the global best fitness improvement is below a chosen threshold. Upon completion, the global best fitness location is the desired output. The PSO pseudo-code is shown in Algorithm \ref{alg:PSO}.

\begin{algorithm}
\caption{Particle Swarm Optimization}\label{alg:PSO}
\begin{algorithmic}[1]
\State Choose initial weights, number of particles, maximum iterations
\State Initialize swarm
\State\hspace{\algorithmicindent} Seed vector created as particle
\State\hspace{\algorithmicindent} Randomly generate remaining particles
\State\hspace{\algorithmicindent} Randomly generate initial velocities
\While {(iterations $<$ maximum iterations) and (improvement $>$ improvement minimum)}
    \State Calculate each particle’s fitness
    \State Update global best fitness and each particle’s best fitness 
    \For {each particle}
        \State Calculate distance between location and particle best fitness
        \State Calculate distance between location and global best fitness
        \State Update particle velocity
        \State Update particle location
    \EndFor
\EndWhile
\State Output global best fitness location
\end{algorithmic}
\end{algorithm}

\subsubsection{Generative Adversarial Network}

Generative adversarial networks are a deep learning technique that pits two neural networks against each other in a game setting. The GAN is composed of a generator and a discriminator. The generator’s task is to learn about the discriminator and thus train to deceive it. The discriminator attempts to discern if its inputs are from the genuine data set or from the adversarial data set. As described in the original work \cite{goodfellow_generative_nodate}, the competition between the two components continues until neither can improve their abilities. In our scenario, it is desired that the generator, once trained, can modify malicious input vectors such that the output is classified as benign by the target classifier. The pseudo-code of the GAN is depicted in Algorithm \ref{alg:GAN}.

\begin{algorithm}
\caption{Generative Adversarial Network}\label{alg:GAN}
\begin{algorithmic}[1]
\For {number of iterations}
    \For {number of generator training steps}
        \State Randomly select rows from benign and malicious traffic vectors
        \State Break malicious vector into mutable and immutable sub-vector
        \State Add noise randomly to malicious mutable sub-vector
        \State Combine malicious noisy mutable and original immutable sub-vectors
        \State Input to generator neural network
        \State Combine mutable output of generator with original immutable sub-vector
        \State Input to discriminator to predict labels
        \State\hspace{\algorithmicindent} If prediction is ‘benign’, feedback label = mutable output of generator
        \State\hspace{\algorithmicindent} If prediction is ‘malicious’, feedback label = benign traffic vector
        \State Train generator using resulting feedback labels
    \EndFor
    \For {number of discriminator training steps}
        \State Randomly select rows from benign traffic vectors
        \State Input benign traffic and generator noisy traffic to discriminator and voting classifier to predict labels 
        \State\hspace{\algorithmicindent} If discriminator label = voting classifier label, feedback = malicious
        \State\hspace{\algorithmicindent} Else, feedback = benign
        \State Train discriminator using resulting feedback labels
    \EndFor
\EndFor
\State Input test vectors to generator to create adversarial vectors
\State Test adversarial vectors using voting classifier 
\end{algorithmic}
\end{algorithm}

Because of the constrained data set, the data input to the generator is divided into mutable and immutable parts like the GA and PSO. Noise is then added to the mutable portion of the data by randomly selecting features and perturbing their values. One contribution that we make in this heuristic is the addition of non-binary features in the GAN’s operations and the consideration of constrained data sets. Other work in this area either does not consider constrained data sets or only use binary features in their operation thus limiting their application. Additionally, due to the nature of this domain, there is no need to search for a minimum perturbation in the data. Unlike the image domain, large changes in the data are not subject to easy recognition by human observers.

\begin{figure}
	\includegraphics[width=15cm]{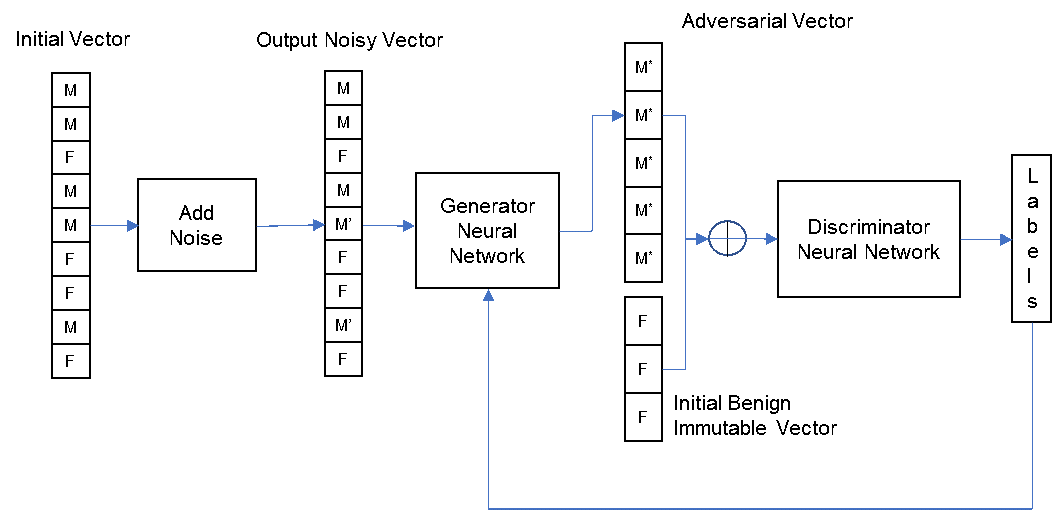}
	\centering
	\caption{GAN generator training operations. M' represents perturbed inputs and M* represents generator modified values.}
	\label{fig:GAN1}
\end{figure}

This noisy vector, containing the noisy mutable and original fixed portions, is then fed to the generator neural network whose output is a vector of the same size as the mutable portion of the original input. This adversarial vector output is then combined with the initial immutable portion of the input vector to form a new adversarial input, which is then fed to the discriminator to generate labels (benign or malicious). These labels are used to improve the neural networks’ predictions. To accomplish our goal, the generator and discriminator are alternately trained using batches of training data. A depiction of the generator’s training is shown in Figure \ref{fig:GAN1}.

\begin{figure}
	\includegraphics[width=15cm]{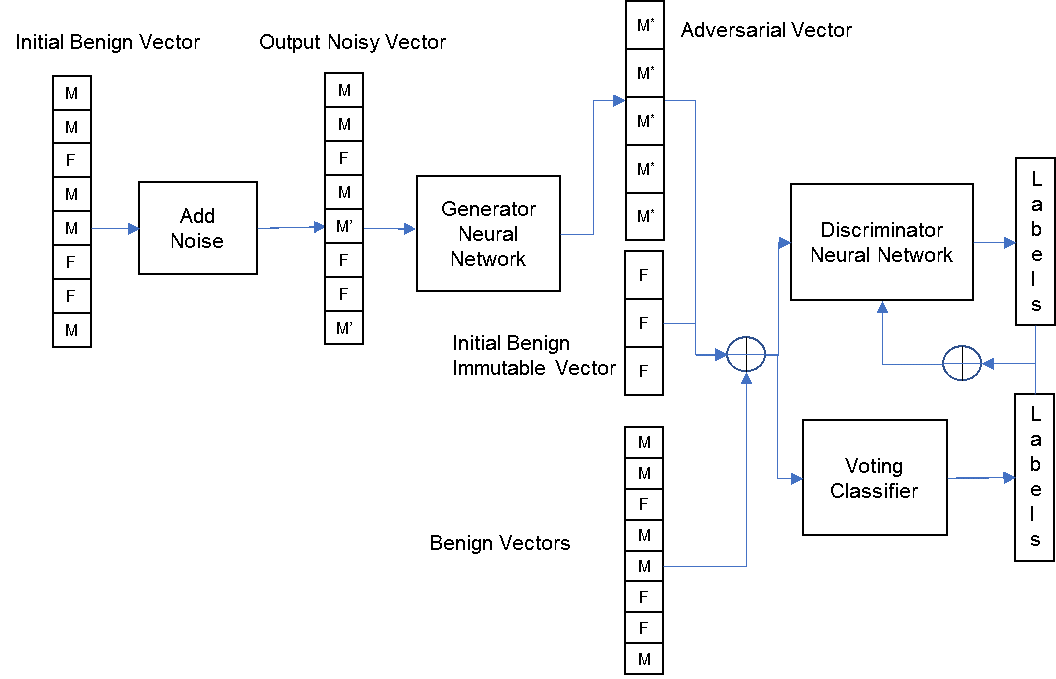}
	\centering
	\caption{GAN discriminator training. M' represents perturbed inputs and M* represents generator modified values.}
	\label{fig:GAN2}
\end{figure}

The discriminator is trained using the complete adversarial vectors from the generator along with benign vectors from the data set. These vectors are then classified by the discriminator and a separate machine learning classifier as benign or malicious. The outputs of the two classifiers are compared in the following way. If both classify an input as malicious, then the feedback to the discriminator is labelled malicious. Otherwise, the classification is benign. A depiction of the discriminator training method is shown in Figure \ref{fig:GAN2}.

\subsubsection{Monte Carlo Simulation}

To create a baseline comparison for the evolutionary computation (GA and PSO) and deep learning (GAN) adversarial example generation techniques, we implemented a MC simulation that randomly perturbs features. The heuristic operates by first randomly selecting the features to perturb from the list of mutable ones. Then, these features are increased in the range $[x_i,1]$. Note that we abide by the same limitations on how the values are modified as in the other three methods. This process is repeated for a fixed number of iterations and the best fitness score is recorded.

\subsection{Computational Experiment}

The perturbation methods based on evolutionary computation, deep learning, and MC simulation were run on both data sets: NSL-KDD \cite{10.5555/1736481.1736489} and UNSW-NB15 \cite{7348942}. Only the vectors labeled as malicious were used in the evaluation of the methods because the ultimate goal is deceiving the classifiers with modified malicious inputs. The NSL-KDD testing set contains $12,828$ malicious vectors while the UNSW-NB15 test set contains $45,328$ malicious vectors. The Scikit-learn package \cite{pedregosa_scikit-learn:_nodate} is used for the development of our code with the exception of the GAN which utilizes Keras models for the neural networks. A baseline classification is developed on the unmodified test set using classifiers similar to those described in the work of \cite{iglesias_analysis_2015}.  The fitness function used for GA and PSO is a Voting Ensemble Classifier formed by soft voting and the following sub-models: Support Vector Machine (SVM), Decision Tree (DT), Naive Bayes (NB), and K-Nearest Neighbors (KNN). The SVM settings are: $gamma=0.01$, $C=220.0$, $tol=0.01$, $probability=True$.  The Naive Bayes model has the default settings, while the Decision Tree algorithm uses the settings: $criterion=``entropy"$, $min\_samples\_split=4$, $min\_samples\_leaf=2$, $max\_depth=20$, $min\_impurity\_decrease=0.1$.  Finally, the KNN algorithm settings are: $n\_neighbors=3$, $algorithm=``auto"$.  
	
The parameters of the GA are a population size of $100$ chromosomes, $1000$ generations maximum, a cross-over rate of $0.25$, a mutation rate of $0.2$, and an early termination criterion of less than $0.01\%$ improvement in 5 generations. The parameters used in the PSO heuristic are $200$ particles, $c_1 = 0.5$ (coefficient for particle local best solution), $c_2 = 0.4$ (coefficient for swarm global best solution), $w = 0.7$ (weight for last velocity), a maximum of $100$ iterations, and an early termination criterion of less than $0.001\%$ improvement. The MC simulation is run with a maximum of $15$ perturbed features per vector and a total of $250$ iterations per vector. The optimal settings on all the algorithms are determined experimentally.
	
The GAN generator is created using a Keras Sequential model with a dense input layer of size $122$ for the NSL-KDD data set and $196$ for the UNSW-NB15 data set. The generator has five additional dense layers of size $29$ for the NSL-KDD data set and $23$ for the UNSW-NB15 data set, and uses a $relu$ activation function for all layers. The model employs a stochastic gradient descent (SGD) optimizer with an $lr$ of $0.01$, a $clipvalue$ of $0.01$ with a loss function of $mean\_squared\_error$. The GAN discriminator is a three-layer dense model with similar input sizes to the generator but with a single output. The initial two layers use a $relu$ activation function while the output layer uses a $sigmoid$ activation function. The model uses a RMSprop optimizer with an $lr$ of $0.01$, a $clipvalue$ of $0.01$, and a $binary\_crossentropy$ loss function. The models are trained for $300$ iterations of $10$ steps each and vector batch size of $50$.   
	
Each of the methods produces a set of perturbed malicious vectors. These vectors are then input into 11 separate classification models to evaluate their performance. Four of the evaluation classifiers are the sub-models used in the GA/PSO voting ensemble. The remaining seven are:  Random Forest (RF), Multi-layer Perceptron (MLP), Gradient Boosting (GB), Logistic Regression (LR), Linear Discriminant Analysis (LDA), Quadratic Discriminant Analysis (QDA), and Bagging (BAG). The models are chosen to cover a broad range of classifier categories.

\section{Results and Discussion}\label{sec:Res}

\subsection{General Results}

The differences in the two data sets used in this work do not allow for comparison between them in terms of commonalities. The attack types used in the sets only overlap in denial-of-service and probing attacks and the techniques in the attacks differ greatly due to the fifteen plus years between their creation. Additionally, the weakness of the NSL-KDD data set with respect to the composition of its training and test sets creates classification differences that are likely less prevalent in the UNSW-NB15 data set. Finally, the features of the two data sets only overlap in five collected fields. Therefore, the features used to identify malicious traffic are mostly incomparable. The discussion of the results by data set below indicates where useful similarities are found between them.  

It should also be noted that no attempt is made to tune the classifiers for the sake of optimizing their performance under any metric. The classifiers are created to form a basis of comparison across a wide range of classifier types and the salient feature is the similarity of the models in their construction.

\subsection{Results from the NSL-KDD Data Set}

Before analyzing the effectiveness of the perturbation methods for adversarial example generation, the $11$ machine learning models are first evaluated for performance (accuracy) using the full NSL-KDD data set. The MLP performed the best with an accuracy of $83.27\%$ with the BAG ($80.20\%$) and LDA ($79.68\%$) models serving second and third best, respectively.

Recall that the PSO, GA, GAN and MC perturbation methods are used to generate adversarial malicious vectors to emulate an evasion attack against the NIDS. To compare the effectiveness of these four adversarial generation techniques, Table \ref{tab:ResNSLKDD} shows the evasion rate of malicious vectors in terms of percentage of attack traffic classified as normal using the $11$ classification models and compared against the unmodified, original vectors.

\begin{table}
  \centering
  \caption{\textbf{NSL-KDD Data Set.} Evasion Rate of Malicious Vectors  (\% Classified Normal)}
    \begin{tabular}{cccccc}
    \textbf{ML Model} & \multicolumn{1}{c}{\textbf{Original Vectors}} & \multicolumn{1}{c}{\textbf{PSO}} & \multicolumn{1}{c}{\textbf{GA}} & \multicolumn{1}{c}{\textbf{GAN}} & \multicolumn{1}{c}{\textbf{MC}} \\
    \midrule
    \textbf{SVM} & 35.78 & 94.37 & 99.95 & 85.85 & 46.41 \\
    \textbf{DT} & 46.89 & 91.75 & 92.37 & 92.60  & 91.13 \\
    \textbf{NB} & 40.87 & 83.00 & 88.27 & 48.06 & 80.39 \\
    \textbf{KNN} & 34.54 & 54.59 & 68.19 & 48.34 & 69.32 \\
    \textbf{RF} & 37.81 & 34.01 & 46.41 & 85.90  & 45.53 \\
    \textbf{MLP} & 26.95 & 44.43 & 77.87 & 86.18 & 38.87 \\
    \textbf{GB} & 32.03 & 1.02  & 6.46  & 66.01 & 33.49 \\
    \textbf{LR} & 37.57 & 14.19 & 24.46 & 33.94 & 36.63 \\
    \textbf{LDA} & 39.66 & 99.99 & 97.18 & 55.19 & 96.19 \\
    \textbf{QDA} & 44.21 & 56.26 & 67.78 & 66.74 & 61.86 \\
    \textbf{BAG} & 28.78 & 54.81 & 56.15 & 61.61 & 41.64 \\
    \midrule
    \textbf{Average} & 36.83 & 57.13 & 65.92 & 66.40 & 58.31 \\
    \end{tabular}%
  \label{tab:ResNSLKDD}%
\end{table}%

Compared to the baseline Monte Carlo perturbation technique, the PSO, GA and GAN performed strictly better for the SVM, DT, MLP, and BAG machine learning models. For the other classification models, there were varying results across the four perturbation methods. When comparing these adversarial generation techniques against the original vectors, the NIDS evasion rate was strictly better for all classifiers except for RF, GB, and LR. This also holds true for the average evasion rate across all classifiers for each adversarial example generation algorithm. The ensemble method classifiers were more resilient on average to adversarial attacks though the GAN had more success against them than the other methods. The evolutionary computation methods (PSO and GA) performed strictly better than the deep learning method (GAN) for the SVM, NB, KNN, and LDA classifiers, whereas the GAN performed better than both PSO and GA for the DT, RF, MLP, GB, LR and BAG classifiers. The former's performance can be attributed partially to the use of most of those models in the evaluation function for the heuristics.  The GAN's superior performance against the remaining classifiers is partially attributed to the learning of its discriminator using the voting classifier and the generator.  Additionally, this suggests that when the classifier under-the-hood of the NIDS uses a tree-based classification model (DT, RF, GB and BAG), then the GAN-based perturbation method is capable of achieving high misclassification rates. 

\begin{figure}
	\includegraphics[width=16cm]{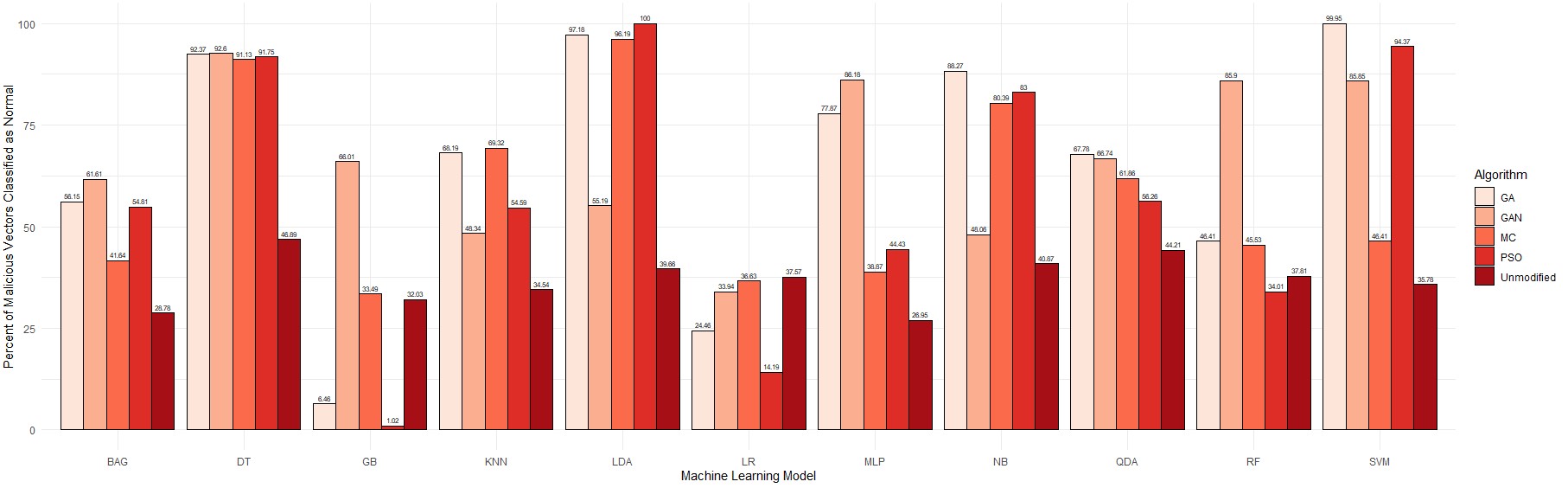}
	\centering
	\caption{Adversarial Example Generation Algorithm Performance Against NSL-KDD Data Set}
	\label{fig:NSLKDD}
\end{figure}

Upon examination of Figure \ref{fig:NSLKDD}, it is evident that the DT classifier has the least variability across the four perturbation methods, with an evasion rate above $90\%$, which is more than double the performance of the original vectors. This holds true for the UNSW-NB15 data set as well. In this experiment, the DT classifier was extremely vulnerable to adversarial examples in general. It is also evident that for QDA and BAG classifiers, the four perturbation methods had limited performance in terms of evasion rate when compared to the original vectors. For the LR classifier, all four perturbation methods performed strictly worse than the unmodified vectors; the results were somewhat mixed for the GB classifier. The best evasion rates by adversarial generation method are depicted in Table \ref{tab:BestFooll}.

\begin{table}
  \centering
  \caption{NSL-KDD Data Set. Best Evasion Rate (\%) by Adversarial Generation Method}
    \begin{tabular}{cccc}
    \multicolumn{1}{c}{\textbf{MC}} & \multicolumn{1}{c}{\textbf{PSO}} & \multicolumn{1}{c}{\textbf{GA}} & \multicolumn{1}{c}{\textbf{GAN}} \\
    \midrule
    96.19 & 99.99 & 99.95 & 92.60 \\
    \end{tabular}%
  \label{tab:BestFooll}%
\end{table}%

\subsection{Results from the UNSW-NB15 Data Set}

For the UNSW-NB15 data set, Table \ref{tab:ResUNSWNB15} highlights the comparative analysis of the four perturbation methods for adversarial generation against the 11 machine learning models representing the NIDS classifier. Again, the performance is represented by the evasion rate of malicious vectors in terms of percentage classified as normal using the classifiers and compared against the original vectors. On average, the GA performed strictly better than the PSO, GAN and MC techniques and greatly fooled the NIDS compared to the original vectors (73.71\% compared to 5.00\% accuracy).

\begin{table}
  \centering
  \caption{\textbf{UNSW-NB15 Data Set.} Evasion Rate of Malicious Vectors  (\% Classified Normal)}
    \begin{tabular}{cccccc}
    \textbf{ML Model} & \multicolumn{1}{c}{\textbf{Original Vectors}} & \multicolumn{1}{c}{\textbf{PSO}} & \multicolumn{1}{c}{\textbf{GA}} & \multicolumn{1}{c}{\textbf{GAN}} & \multicolumn{1}{c}{\textbf{MC}} \\
    \midrule
    \textbf{SVM} & 0.37  & 96.85 & 98.88 & 22.04 & 77.67 \\
    \textbf{DT} & 0.14  & 96.85 & 100.00 & 95.43 & 99.53 \\
    \textbf{NB} & 33.06 & 83.66 & 75.56 & 32.98 & 32.90 \\
    \textbf{KNN} & 3.82  & 23.56 & 65.34 & 4.19  & 10.64 \\
    \textbf{RF} & 0.64  & 9.55  & 31.28 & 30.61 & 46.04 \\
    \textbf{MLP} & 0.63  & 11.74 & 39.36 & 53.90 & 55.10 \\
    \textbf{GB} & 0.55  & 98.06 & 99.74 & 99.42 & 99.39 \\
    \textbf{LR} & 2.51  & 86.96 & 91.25 & 71.54 & 88.77 \\
    \textbf{LDA} & 0.21  & 85.68 & 84.89 & 7.21  & 55.09 \\
    \textbf{QDA} & 10.18 & 4.77  & 24.82 & 31.37 & 30.42 \\
    \textbf{BAG} & 2.91  & 97.83 & 99.65 & 99.61 & 99.30 \\
    \midrule
    \textbf{Average} & 5.00  & 63.23 & 73.71 & 49.85 & 63.17 \\
    \end{tabular}%
  \label{tab:ResUNSWNB15}%
\end{table}%

Compared to the MC technique as the baseline perturbation method, PSO, GA and GAN performed strictly better for the NB  machine learning models, and the results were greatly mixed for the other classifiers representing the NIDS. When comparing the four adversarial generation techniques against the original vectors, the NIDS evasion rate was strictly better for all classifiers except NB and QDA suggesting that modern traffic features are more susceptible to adversarial manipulation than the older traffic. The evolutionary computation methods (PSO and GA) performed strictly better than the deep learning method (GAN) for the SVM, DT, NB, KNN, LR, and LDA classifiers, whereas the GAN performed strictly better than both PSO and GA for the MLP and QDA. This seems to suggest that when the classifier under-the-hood of the NIDS uses a neural network based classification model (MLP), then the GAN-based perturbation method can more easily fool the NIDS; although, the MC technique did an overall better job of evading  for this machine learning model. 

\begin{figure}
	\includegraphics[width=16cm]{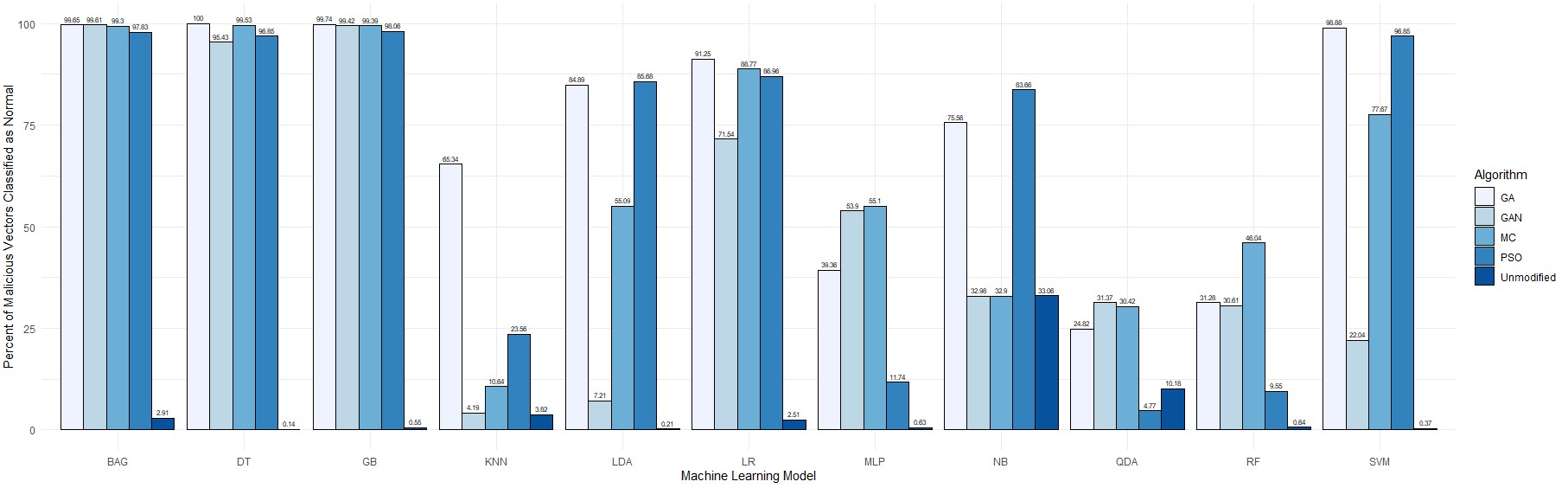}
	\centering
	\caption{Adversarial Example Generation Algorithm Performance Against UNSW-NB15 Data Set}
	\label{fig:UNSWNB15}
\end{figure}

Upon examination of Figure \ref{fig:UNSWNB15}, it is evident that the BAG, DT and GB classifiers had the least variability across the four perturbation methods, with an evasion rate above 95\%. This suggests that each of the adversarial example generation techniques are nearly equally good at fooling NIDS that use tree-based classifiers; this also suggests that these types of machine learning models are easily fooled. This is similar to the RF classifier, which is also tree-based, but the overall performance is much less for the four perturbation methods. For the NB classifier, the GAN and MC performed strictly worse than the unmodified vectors, which was not the case for the evolutionary computation methods (PSO and GA) that performed strictly better (over double). The best evasion rates by adversarial generation method are depicted in Table \ref{tab:BestFool}.

\begin{table}
  \centering
  \caption{UNSW-NB15 Data Set. Best Evasion Rate (\%) by Adversarial Generation Method}
    \begin{tabular}{cccc}
    \multicolumn{1}{c}{\textbf{MC}} & \multicolumn{1}{c}{\textbf{PSO}} & \multicolumn{1}{c}{\textbf{GA}} & \multicolumn{1}{c}{\textbf{GAN}} \\
    \midrule
    99.53 & 98.06 & 100.00 & 99.61 \\
    \end{tabular}%
  \label{tab:BestFool}%
\end{table}%

\section{Conclusion}\label{sec:Conc}

It is widely believed nowadays that no domain is immune to adversarial machine learning. With the purpose of validating the hypothesis that constrained domains are at least as vulnerable to adversarial attacks as unconstrained ones, this paper investigated the effects of creating perturbations using techniques from evolutionary computation (PSO and GA) and deep learning (GAN). The main goal of our computational experimentation was to tweak malicious traffic in order to evade detection by NIDS, i.e. to ``trick" the ML systems to classify such traffic as normal.

To the best of our knowledge, this work is the first to incorporate evolutionary algorithms in the realm of adversarial machine learning in the network intrusion detection setting. Throughout our results, it was evident that the data vectors created via the PSO, GA, and GAN methods achieved high misclassification rates against a handful of machine learning classifiers. We highlight some key takeaways below.

First, the support vector machine (SVM) and the decision tree (DT) classifiers were the most vulnerable against adversarial examples created by evolutionary algorithms and generative adversarial networks ($>90\%$ evasion rate). This leads us to recommend refraining from using them in NIDS and in mission-critical tasks in general. Second, in the network traffic data, it is not quite realistic to assume that a potential adversary has the capability to alter each and every feature of the traffic. This is a crucial assumption that distinguishes between crafting adversarial examples in unconstrained domains versus ones in constrained domains. In our work, we rely on subject matter expertise to filter the features that can be modified by an attacker without breaking the functionality of the network traffic. It is worth mentioning that our strategy differs from that adopted in \cite{sheatsley_adversarial_nodate}, where the choice of modifiable features does not seem to preserve the network functionality. Third, our results show that the same set of adversarial examples that managed to fool one machine learning classifier also succeeded at fooling several others. A particular instance of such observation was the performance of the adversarial examples generated by the PSO algorithm in the UNSW-NB15 data set ($98.06\%$ against GB, $85.68\%$ against LDA, and $97.83\%$ against BAG). This observation can be considered as additional evidence to the transferability phenomenon first alluded to in \cite{papernot2016transferability} within the image recognition setting and in \cite{sheatsley_adversarial_nodate} within the network intrusion detection setting.

In the future, we aim to analyze the internal operations of the ML models used in this paper. It is clear that all of them are vulnerable to adversarial perturbations, but it not clear why some of them are more robust than the others in the NIDS setting. We hope that a systematic study of their mechanism would shed some light on their robustness and explain their sensitivity to small input alterations.

\section*{Disclaimer}
The views expressed in this paper are those of the authors and do not reflect the official policy or position of the United States Military Academy, the United States Army, the Department of Defense, or the United States Government.

\section*{Acknowledgments}
This work was partially funded by the Army Research Laboratory (ARL) through the Mathematical Sciences Center, Department of Mathematical Sciences, U.S. Military Academy, West Point, NY.  The authors would like to thank the U.S. Army's Engineering Research and Development Center (ERDC) for use of their compute resources in the course of our work.

\printbibliography

\end{document}